\documentclass[conference]{IEEEtran}
\usepackage[T1]{fontenc}
\usepackage[utf8]{inputenc}
\usepackage{cite}
\usepackage{graphicx}
\usepackage{booktabs}
\usepackage{amsmath}
\usepackage{amssymb}
\usepackage{array}
\usepackage{url}
\usepackage{multirow}
\usepackage{longtable}
\usepackage[caption=false,font=footnotesize]{subfig}
\usepackage{tikz}
\usetikzlibrary{arrows.meta,positioning,shapes.geometric,fit,calc}
\usepackage{balance}

\graphicspath{{figures/}}

\title{APEX: Agent Payment Execution with Policy for Autonomous Agent API Access}
\author{
\IEEEauthorblockN{Mohd Safwan Uddin, Mohammed Mouzam, Mohammed Imran, Syed Badar Uddin Faizan}
\IEEEauthorblockA{Email: safwanuddin405@gmail.com, mouzam.cse@gmail.com, imranmohd.cse1@gmail.com, syed7013117@gmail.com}
}
\begin{document}
\maketitle

\begin{abstract}
Autonomous agents are moving beyond simple retrieval tasks to become economic actors that invoke APIs, sequence workflows, and make real-time decisions.
As this shift accelerates, API providers need request-level monetization with programmatic spend governance.
The HTTP 402 protocol addresses this by treating payment as a first-class protocol event, but most implementations rely on cryptocurrency rails.
In many deployment contexts—especially countries with strong real-time fiat systems like UPI—this assumption is misaligned with regulatory and infrastructure realities.

We present APEX, an implementation-complete research system that adapts HTTP 402-style payment gating to UPI-like fiat workflows while preserving policy-governed spend control, tokenized access verification, and replay resistance.
We implement a challenge-settle-consume lifecycle with HMAC-signed short-lived tokens, idempotent settlement handling, and policy-aware payment approval.
The system uses FastAPI, SQLite, and Python standard libraries, making it transparent, inspectable, and reproducible.

We evaluate APEX across three baselines (no policy, payment without policy, payment with policy) and six scenarios (normal, overspending, replay attack, invalid token, token expiry, idempotency) using sample sizes 2-4x larger than initial experiments (N=20-40 per scenario).
Results show that policy enforcement reduces total spending by 27.3\% (\$550 to \$400) while maintaining 52.8\% success rate for legitimate requests.
Security mechanisms achieve 100\% block rate for both replay attacks (20/20 blocked) and invalid tokens (20/20 blocked) with low latency overhead (19.6ms average).
Multiple trial runs show low variance (±2.7-8.9ms standard deviation) across scenarios, demonstrating high reproducibility with 95\% confidence intervals.
Payment gating introduces 86.9ms average latency overhead compared to 8.0ms baseline, representing a 10.9x slowdown that remains acceptable for controlled agent payment workflows in research contexts.

The primary contribution is a controlled agent-payment infrastructure and reference architecture that demonstrates how agentic access monetization can be adapted to fiat systems without discarding security and policy guarantees.
\end{abstract}


\section{Introduction}
Autonomous agents are moving beyond simple retrieval tasks.
They now invoke APIs, sequence multi-step workflows, and make bounded execution decisions in real time~\cite{react,toolformer,autogpt}.
As this shift accelerates, API providers increasingly require request-level monetization and spend governance that can be enforced programmatically, not manually.

The emerging HTTP 402-oriented ecosystem addresses this need by treating payment as a first-class protocol event, not an afterthought in monthly invoicing pipelines.
In this interaction model, an unpaid request receives a machine-readable payment challenge, and the client retries with proof of payment.
This pattern is attractive for agent-to-agent markets, where low-friction, small-value, high-frequency transactions are common.

Yet most practical demonstrations around this direction are tightly linked to blockchain rails.
In many deployment contexts, that assumption is misaligned with regulatory, user, and infrastructure realities.
In countries like India where UPI processes billions of transactions monthly~\cite{bisupi,upigrowth}, agents need to work with fiat systems, not crypto.
The research question is therefore direct: can we preserve the architectural strengths of 402-style access control while grounding settlement semantics in fiat-like flows and enforcing policy at payment time?

This paper answers that question through APEX, an implementation-complete reference architecture built for controlled experimentation.
The system is intentionally scoped to protocol behavior, policy enforcement, security invariants, structured logging, and repeatable scenario-based measurements rather than full banking integration.

APEX contributes five practical elements.
First, it adapts a challenge-based 402 interaction to a UPI-like payment abstraction, keeping the developer-facing API simple.
Second, it enforces spend control using request limits and daily budget policy checks as the primary control boundary.
Third, it secures post-payment access with signed, expiring, single-use tokens.
Fourth, it supports idempotent settlement to reduce duplicate side effects under retries.
Fifth, it includes an experiment harness that directly compares baseline behavior under normal and adversarial patterns.

\subsection{Key Contributions}
We make four concrete contributions aimed at reproducible systems research.
\begin{enumerate}
\item \textbf{Fiat-oriented protocol adaptation:}
An HTTP 402-style challenge-settle-retry pattern mapped to UPI-like payment semantics for agent-executed API access.
\item \textbf{Policy-governed payment control:}
Deterministic per-request and daily budget enforcement integrated directly into settlement decisions.
\item \textbf{Security-complete token lifecycle:}
Signed, expiring, single-use tokens with replay rejection and idempotent settlement behavior under retries.
\item \textbf{End-to-end reproducibility:}
Structured logs, scenario baselines, and machine-readable outputs that directly support publication tables and figures.
\end{enumerate}

A major design goal was a small dependency surface.
The implementation uses FastAPI, SQLite, and standard Python libraries for cryptography, serialization, and timing, matching constraints common in reproducible systems research.
This choice intentionally favors interpretability over feature breadth.

The remainder of the paper is organized as follows.
Section II reviews related work in HTTP 402, micropayments, agent systems, and payment policy control.
Section III defines the problem, assumptions, and threat model.
Section IV describes the APEX architecture and lifecycle.
Section V defines system constraints and latency models.
Section VI details implementation.
Section VII documents experimental design.
Section VIII reports results.
Section IX discusses implications and limitations.
Section X outlines future work, and Section XI concludes.

\section{Related Work}
\subsection{HTTP 402 and Agent-Native Payment Flows}
The x402 protocol extends HTTP 402 to enable internet-native API monetization, where payment is integrated directly into request handling~\cite{x402}.
This positions 402 not merely as a legacy status code, but as a practical trigger for machine-interpretable payment challenges.
Extensions like A402 explore atomic service channels and address latency concerns in payment settlement~\cite{a402}.
Recent work examines how 402-style semantics could enable autonomous transactions between software agents~\cite{a2a402,microeconagent,trustfabric}.

These works establish architectural motivation, but most available artifacts remain focused on crypto rails.
The fiat adaptation problem remains under-explored, especially for systems intended to interoperate with UPI-like abstractions.
APEX addresses that adaptation gap with a strict, inspectable implementation scope and an explicit policy boundary, where spend governance is the decision point that distinguishes controlled agent monetization from simple payment transport.

\subsection{Micropayments and Settlement Infrastructure}
Micropayment literature emphasizes fee efficiency, settlement reliability, and machine-scale throughput.
Systems such as MicroCash and Lightning-based designs discuss probabilistic, off-chain, or channel-based mechanisms for very small transactions~\cite{microcash,feeless,lniot,lniotfgcs}.
These studies are valuable for performance intuition, but deployment assumptions differ from fiat-led payment ecosystems.

In fiat systems, especially near-real-time rails, compliance, identity, and institutional integration constraints dominate design decisions.
A practical system in this setting must expose where protocol-level ideas transfer directly, and where architecture must diverge.

\subsection{AI Agents and API Execution}
Agent frameworks continue to evolve from reasoning-only systems toward tool-using systems that execute side effects.
ReAct, Toolformer, and autonomous agent frameworks illustrate this progression~\cite{react,toolformer,autogpt}.
As execution autonomy increases, spend and access controls become mandatory, not optional.
Survey work on web-capable agents also highlights risks tied to unconstrained tool invocation and external action loops~\cite{webagents}.

APEX situates itself in this execution-centric framing.
Its value is not in generalized cognition, but in disciplined payment-mediated API access for controlled agent behavior.

\subsection{UPI and Real-Time Fiat Systems}
UPI literature and policy analyses document large-scale adoption, low settlement latency, and broad user accessibility, making UPI-like rails strategically relevant for research on practical agentic payments~\cite{bisupi,upigrowth,rtppaper}.
While these sources do not define agent protocols, they provide the operational context motivating fiat-oriented implementations.

\subsection{Policy Enforcement and API Control}
Policy-governed API execution is established in rate limiting, budget governance, and adaptive gateway design~\cite{apiratelimit,aigateway,gcpbudget}.
APEX draws from this line of work by enforcing bounded per-request and cumulative spend policies at payment-time, with deterministic reject behavior under violations.

\subsection{Research Gap}
The current landscape still lacks open, reproducible systems that combine:
(1) a 402-like challenge flow,
(2) fiat-oriented settlement semantics,
(3) explicit policy controls,
(4) tokenized replay-resistant verification,
and
(5) benchmark-style scenario experiments with exportable metrics.
APEX is designed as a direct contribution to this specific gap.

Table~\ref{tab:x402-compare} positions APEX against representative x402-style systems at a high level.

\begin{table}[t]
\caption{Comparison with Existing x402-based Systems}
\label{tab:x402-compare}
\centering
\begin{tabular}{lcc}
\toprule
\textbf{Feature} & \textbf{x402} & \textbf{APEX} \\
\midrule
Fiat support & No & Yes \\
Policy control & Limited & Yes \\
Replay protection & Partial & Strong \\
Experimental validation & Limited & Yes \\
\bottomrule
\end{tabular}
\end{table}

\section{Problem Statement and Scope}
\subsection{Core Problem}
Given an API endpoint that should only return protected data after successful payment, we require a machine-executable protocol that:

\begin{enumerate}
\item challenges unpaid requests with structured payment details,
\item accepts a payment settlement attempt,
\item validates payment proof on subsequent access,
\item enforces spend policy constraints,
\item prevents replay and token forgery abuse,
\item and provides measurable logs for empirical evaluation.
\end{enumerate}

The problem is constrained to request-level payments, not subscriptions, not credit models, and not invoice post-processing.

\subsection{Design Objectives}
APEX was developed under five explicit objectives.

\begin{enumerate}
\item \textbf{Protocol clarity:}
A simple and explicit challenge-settle-consume state progression.
\item \textbf{Policy determinism:}
Hard rejection when request amount or daily budget constraints are violated.
\item \textbf{Security by default:}
Short-lived signed tokens, single-use semantics, and replay rejection.
\item \textbf{Operational observability:}
Append-only structured logs with status, reason, latency, and endpoint context.
\item \textbf{Experimental reproducibility:}
Built-in scenario runner that exports comparable metrics for paper tables and figures.
\end{enumerate}

\subsection{Out of Scope}
The following are intentionally excluded from this implementation scope.

\begin{enumerate}
\item Real bank settlement integration, KYC, and financial compliance workflows.
\item Horizontal distributed consensus or multi-node fault tolerance.
\item End-user UI/UX design and payment confirmation interfaces.
\item Production-grade secrets management and HSM-backed signing.
\item Formal verification of protocol implementation.
\end{enumerate}

These exclusions preserve focus on architectural and experimental clarity.

\section{System Model and Threat Model}
\subsection{System Entities}
APEX models five entities.

\begin{enumerate}
\item \textbf{Client Agent:}
Calls the protected endpoint, parses challenge details, and performs settlement attempts.
\item \textbf{Protected API:}
Exposes \texttt{/data} and returns either a challenge or protected payload.
\item \textbf{Payment API:}
Exposes \texttt{/pay}, performs policy checks, settles payment state, and returns verification token.
\item \textbf{Ledger Store:}
SQLite table recording state, tokens, amount, and idempotency metadata.
\item \textbf{Policy Engine:}
Evaluates per-request and cumulative spend constraints.
\end{enumerate}

\subsection{Adversary Capabilities}
The adversary can:

\begin{enumerate}
\item call endpoints without payment,
\item replay a previously consumed token,
\item submit malformed or forged tokens,
\item attempt overspending through repeated calls,
\item and trigger duplicate settlement requests.
\end{enumerate}

The adversary cannot:

\begin{enumerate}
\item compromise server-side secret key storage,
\item alter server code at runtime,
\item or tamper with database files directly.
\end{enumerate}

\subsection{Security Goals}
APEX aims to satisfy four goals.

\begin{enumerate}
\item \textbf{G1 - Access control:}
Protected data is returned only after valid settlement evidence.
\item \textbf{G2 - Integrity:}
Forged tokens are rejected.
\item \textbf{G3 - Replay resistance:}
Consumed tokens cannot grant repeated access.
\item \textbf{G4 - Policy enforcement:}
Requests violating spend constraints are blocked deterministically.
\end{enumerate}

\subsection{Failure Categories}
Observed outcomes are grouped into:

\begin{enumerate}
\item \texttt{success}
\item \texttt{blocked}
\item \texttt{failed}
\end{enumerate}

This triad is used consistently across API responses, structured logs, and experiment summaries.

\begin{figure}[t]
\centering
\begin{tikzpicture}[
font=\footnotesize,
box/.style={draw, rounded corners=2pt, align=center, minimum height=0.62cm},
arr/.style={-{Latex[length=2mm]}, thick}
]
\node[box, fill=red!12, draw=red!60!black, minimum width=2.35cm, minimum height=0.74cm] (attacker) at (0,0) {Attacker Node};

\node[box, fill=blue!8, draw=blue!55!black, minimum width=3.45cm] (api) at (5.1,0.95) {API Server\\(\texttt{/data}, \texttt{/pay})};
\node[box, fill=purple!10, draw=purple!55!black, minimum width=3.45cm] (token) at (5.1,0.00) {Token Service\\(verify + expiry + consume)};
\node[box, fill=green!10, draw=green!45!black, minimum width=3.45cm] (policy) at (5.1,-0.95) {Policy Engine\\(max/request + daily budget)};

\node[draw, dashed, rounded corners=2pt, fit=(api)(token)(policy), inner sep=7pt, label={[font=\scriptsize]above:Defended Backend Surface}] (backend) {};

\draw[arr, red!70!black] (attacker.east) to[out=30, in=180] node[pos=0.42, above, font=\scriptsize, fill=white, inner sep=1pt] {A1} (api.west);
\draw[arr, red!70!black] (attacker.east) -- node[pos=0.42, above, font=\scriptsize, fill=white, inner sep=1pt] {A2} (token.west);
\draw[arr, red!70!black] (attacker.east) to[out=-30, in=180] node[pos=0.42, below, font=\scriptsize, fill=white, inner sep=1pt] {A3} (policy.west);

\draw[arr, blue!70!black] (api.south) -- (token.north);
\draw[arr, blue!70!black] (token.south) -- (policy.north);

\node[anchor=west, font=\scriptsize, text=red!70!black] at (-1.35,-1.55) {A1: invalid token};
\node[anchor=west, font=\scriptsize, text=red!70!black] at (-1.35,-1.85) {A2: replay attempt};
\node[anchor=west, font=\scriptsize, text=red!70!black] at (-1.35,-2.15) {A3: overspending flood};
\end{tikzpicture}
\caption{Threat Model and Attack Surfaces}
\label{fig:threat-model-apex}
\end{figure}
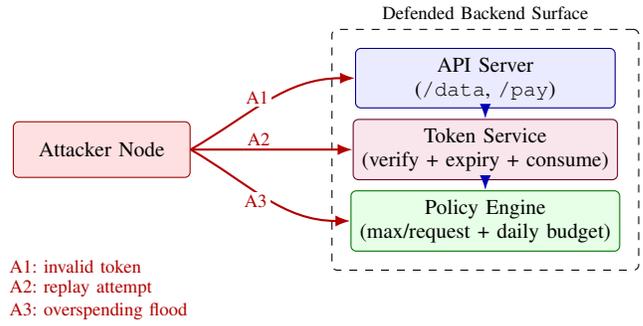

Figure~\ref{fig:threat-model-apex} summarizes the security boundary and attack surfaces considered in this work.

\section{APEX Architecture}
\subsection{Endpoint Overview}
The architecture centers around two externally invoked endpoints and one auxiliary reset endpoint used for controlled experiments.

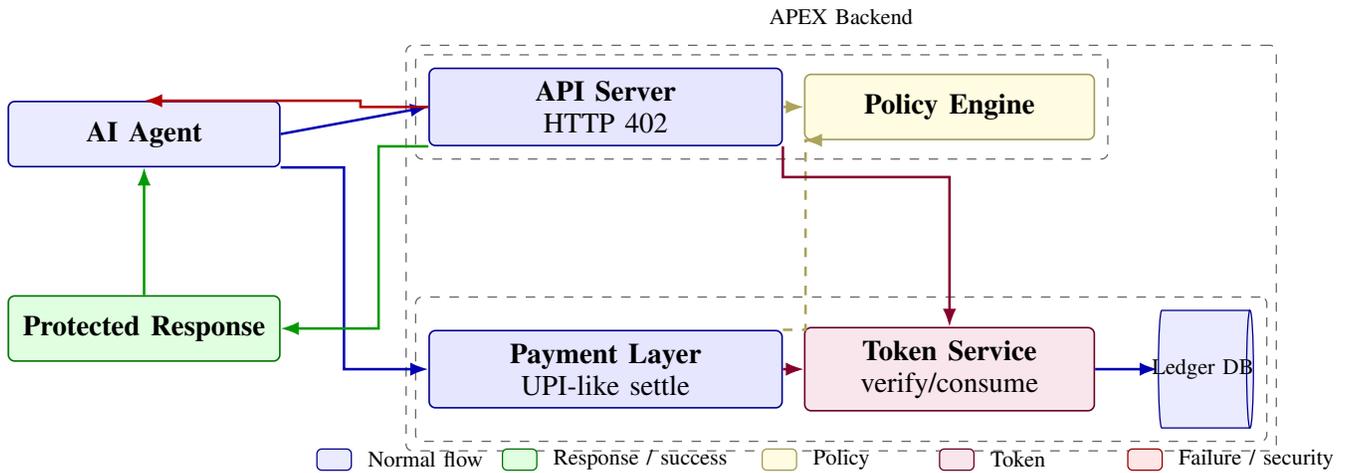
\begin{figure*}[!t]
\centering
\resizebox{0.98\textwidth}{!}{%
\begin{tikzpicture}[
font=\small,
box/.style={draw, rounded corners=2pt, align=center, minimum height=0.72cm, inner sep=4pt, line width=0.5pt},
arr/.style={-{Latex[length=2mm]}, line width=0.8pt},
policyarr/.style={-{Latex[length=2mm]}, line width=0.8pt, dashed}
]
\node[box, fill=blue!8, draw=blue!55!black, minimum width=3.0cm] (agent) at (0,1.45) {\textbf{AI Agent}};
\node[box, fill=green!12, draw=green!45!black, minimum width=3.0cm] (response) at (0,-0.70) {\textbf{Protected Response}};

\node[box, fill=blue!10, draw=blue!55!black, minimum width=3.9cm] (api) at (5.1,1.75) {\textbf{API Server}\\HTTP 402};
\node[box, fill=yellow!14, draw=yellow!55!black, minimum width=3.2cm] (policy) at (8.9,1.75) {\textbf{Policy Engine}};

\node[box, fill=blue!10, draw=blue!55!black, minimum width=3.9cm] (payment) at (5.1,-1.15) {\textbf{Payment Layer}\\UPI-like settle};
\node[box, fill=purple!10, draw=purple!55!black, minimum width=3.2cm] (token) at (8.9,-1.15) {\textbf{Token Service}\\verify/consume};

\node[cylinder, draw=blue!55!black, fill=blue!8, aspect=0.32, minimum height=1.05cm, minimum width=1.30cm] (ledger) at (11.7,-1.15) {};
\node[font=\scriptsize] at (ledger.center) {Ledger DB};

\node[draw=black!60, dashed, rounded corners=2pt, fit=(api)(policy)(payment)(token)(ledger), inner sep=7pt] (backend) {};
\node[draw=black!60, dashed, rounded corners=2pt, fit=(api)(policy), inner sep=4pt] (layerA) {};
\node[draw=black!60, dashed, rounded corners=2pt, fit=(payment)(token)(ledger), inner sep=4pt] (layerB) {};
\node[font=\scriptsize, anchor=south, yshift=3pt] at (backend.north) {APEX Backend};

\draw[arr, blue!70!black] (agent.east) -- (api.west);
\draw[arr, red!70!black] (api.west) -- ++(-0.75,0) |- (agent.north);
\draw[arr, blue!70!black] (agent.south east) -- ++(0.70,0) |- (payment.west);

\draw[policyarr, yellow!60!black] (api.east) -- (policy.west);
\draw[policyarr, yellow!60!black] (payment.north east) -- ++(0.25,0) |- (policy.south west);

\draw[arr, purple!70!black] (payment.east) -- (token.west);
\draw[arr, purple!70!black] (api.south east) -- ++(0,-0.34) -| (token.north);
\draw[arr, blue!70!black] (token.east) -- (ledger.west);

\draw[arr, green!60!black] (api.south west) -- ++(-0.55,0) |- (response.east);
\draw[arr, green!60!black] (response.north) -- (agent.south);

% Compact legend
\node[draw=blue!55!black, fill=blue!8, rounded corners=1.5pt, minimum width=0.38cm, minimum height=0.18cm] at (2.1,-2.15) {};
\node[anchor=west, font=\scriptsize] at (2.35,-2.15) {Normal flow};
\node[draw=green!45!black, fill=green!12, rounded corners=1.5pt, minimum width=0.38cm, minimum height=0.18cm] at (4.15,-2.15) {};
\node[anchor=west, font=\scriptsize] at (4.40,-2.15) {Response / success};
\node[draw=yellow!55!black, fill=yellow!14, rounded corners=1.5pt, minimum width=0.38cm, minimum height=0.18cm] at (7.02,-2.15) {};
\node[anchor=west, font=\scriptsize] at (7.27,-2.15) {Policy};
\node[draw=purple!55!black, fill=purple!10, rounded corners=1.5pt, minimum width=0.38cm, minimum height=0.18cm] at (8.98,-2.15) {};
\node[anchor=west, font=\scriptsize] at (9.23,-2.15) {Token};
\node[draw=red!60!black, fill=red!10, rounded corners=1.5pt, minimum width=0.38cm, minimum height=0.18cm] at (11.06,-2.15) {};
\node[anchor=west, font=\scriptsize] at (11.31,-2.15) {Failure / security};
\end{tikzpicture}%
}
\caption{System Architecture of APEX}
\label{fig:system-architecture-apex}
\end{figure*}

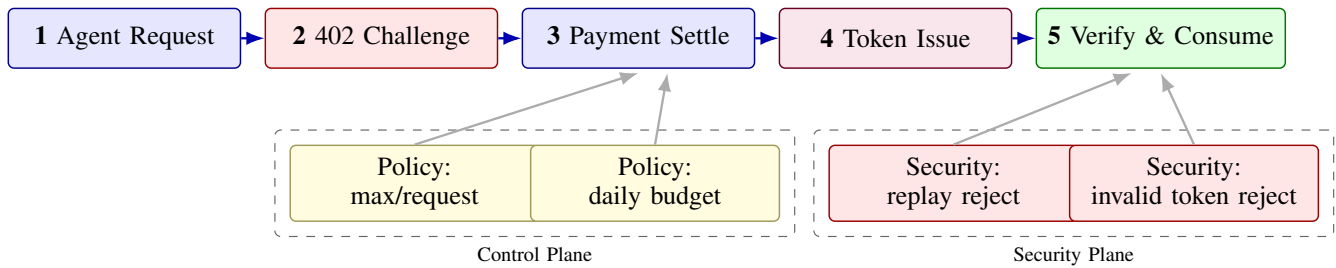
\begin{figure*}[t]
\centering
\resizebox{0.98\textwidth}{!}{%
\begin{tikzpicture}[
font=\small,
box/.style={draw, rounded corners=2pt, align=center, minimum height=0.72cm, inner sep=4pt, line width=0.5pt},
arr/.style={-{Latex[length=2.2mm]}, line width=0.8pt}
]
\node[box, fill=blue!10, draw=blue!55!black, minimum width=2.8cm] (a1) at (0,2.0) {\textbf{1} Agent Request};
\node[box, fill=red!10, draw=red!60!black, minimum width=2.8cm] (a2) at (3.1,2.0) {\textbf{2} 402 Challenge};
\node[box, fill=blue!10, draw=blue!55!black, minimum width=2.8cm] (a3) at (6.2,2.0) {\textbf{3} Payment Settle};
\node[box, fill=purple!10, draw=purple!55!black, minimum width=2.8cm] (a4) at (9.3,2.0) {\textbf{4} Token Issue};
\node[box, fill=green!12, draw=green!45!black, minimum width=3.0cm] (a5) at (12.5,2.0) {\textbf{5} Verify \& Consume};

\node[box, fill=yellow!15, draw=yellow!55!black, minimum width=3.0cm] (p1) at (3.5,0.25) {Policy:\\max/request};
\node[box, fill=yellow!15, draw=yellow!55!black, minimum width=3.0cm] (p2) at (6.4,0.25) {Policy:\\daily budget};
\node[box, fill=red!10, draw=red!60!black, minimum width=3.0cm] (s1) at (10.0,0.25) {Security:\\replay reject};
\node[box, fill=red!10, draw=red!60!black, minimum width=3.0cm] (s2) at (12.9,0.25) {Security:\\invalid token reject};

\node[draw=black!60, dashed, rounded corners=2pt, fit=(p1)(p2), inner sep=5pt] (cp) {};
\node[draw=black!60, dashed, rounded corners=2pt, fit=(s1)(s2), inner sep=5pt] (sp) {};
\node[font=\scriptsize, anchor=north] at (cp.south) {Control Plane};
\node[font=\scriptsize, anchor=north] at (sp.south) {Security Plane};

\draw[arr, blue!70!black] (a1.east) -- (a2.west);
\draw[arr, blue!70!black] (a2.east) -- (a3.west);
\draw[arr, blue!70!black] (a3.east) -- (a4.west);
\draw[arr, blue!70!black] (a4.east) -- (a5.west);

\draw[arr, gray!65] (p1.north) -- ($(a3.south)+(0,-0.05)$);
\draw[arr, gray!65] (p2.north) -- ($(a3.south)+(0.35,-0.05)$);
\draw[arr, gray!65] (s1.north) -- ($(a5.south)+(-0.35,-0.05)$);
\draw[arr, gray!65] (s2.north) -- ($(a5.south)+(0,-0.05)$);
\end{tikzpicture}
}
\caption{End-to-End APEX System Overview}
\label{fig:master-overview-apex}
\end{figure*}

Figure~\ref{fig:system-architecture-apex} presents the end-to-end APEX architecture, including challenge, policy, settlement, verification, and response components.
Figure~\ref{fig:master-overview-apex} provides a single integrated control, security, and payment-flow view for quick system understanding.

\begin{enumerate}
\item \textbf{GET /data}
\begin{enumerate}
\item If baseline is \texttt{no\_policy}, returns protected data directly.
\item If no payment token is provided, creates challenge record and returns HTTP 402 with \texttt{ref\_id} and amount.
\item If token is provided, verifies signature and expiry, then attempts token consumption.
\item Returns data only if token is valid and consumable.
\end{enumerate}

\item \textbf{POST /pay}
\begin{enumerate}
\item Receives \texttt{ref\_id}, \texttt{amount}, \texttt{baseline}, and optional idempotency key.
\item Evaluates policy according to baseline.
\item Issues signed token with expiry.
\item Settles payment state transactionally.
\item Returns token and state metadata on success.
\end{enumerate}

\item \textbf{POST /reset}
\begin{enumerate}
\item Clears ledger table for reproducible scenario execution.
\end{enumerate}
\end{enumerate}

\subsection{State Machine}
Each request reference moves through a strict state sequence:

\begin{enumerate}
\item \textbf{CHALLENGED}:
challenge created at unpaid \texttt{/data} request.
\item \textbf{INITIATED}:
settlement process started.
\item \textbf{SETTLED}:
payment accepted, token attached, idempotency key recorded.
\item \textbf{CONSUMED}:
first valid token use grants data and consumes entitlement.
\end{enumerate}

Transitions are implemented with SQLite transactions using \texttt{BEGIN IMMEDIATE} to reduce race risks in single-node operation.

\subsection{Data Model}
The payment ledger stores:

\begin{enumerate}
\item \texttt{ref\_id} (primary key)
\item \texttt{amount}
\item \texttt{created\_at}
\item \texttt{state}
\item \texttt{token}
\item \texttt{token\_expiry}
\item \texttt{consumed\_at}
\item \texttt{idempotency\_key}
\end{enumerate}

Indexes are maintained for \texttt{state} and \texttt{token} to support lookup efficiency in the system.

\subsection{Baseline Modes}
APEX explicitly supports three comparative baselines.

\begin{enumerate}
\item \texttt{no\_policy}:
No payment gating, direct access path, no spend control.
\item \texttt{payment\_no\_policy}:
Payment challenge and token verification enabled, policy checks disabled.
\item \texttt{payment\_with\_policy}:
Full payment and policy controls enabled.
\end{enumerate}

These modes are selected per request to enable side-by-side controlled experiments under identical runtime stack.

\section{Protocol Walkthrough}
\subsection{Normal Access Sequence}
The standard sequence is:

\begin{enumerate}
\item Agent calls \texttt{GET /data} with baseline \texttt{payment\_with\_policy}.
\item Server responds \texttt{402} with challenge payload including \texttt{ref\_id} and amount.
\item Agent calls \texttt{POST /pay} with challenge values.
\item Policy is evaluated.
\item Token is issued and settlement state becomes \texttt{SETTLED}.
\item Agent retries \texttt{GET /data} with \texttt{x-payment-token}.
\item Token is verified and consumed.
\item Protected content is returned.
\end{enumerate}

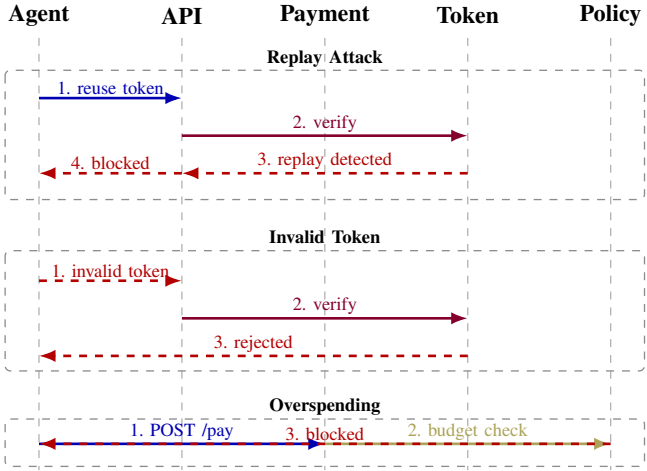
\begin{figure}[t]
\centering
\begin{tikzpicture}[
font=\small,
lane/.style={draw=black!30, dashed, line width=0.5pt},
msg/.style={-{Latex[length=2.2mm]}, line width=1.0pt},
fail/.style={-{Latex[length=2.2mm]}, line width=1.0pt, red!70!black, dashed},
normal/.style={msg, blue!70!black},
token/.style={msg, purple!70!black},
policy/.style={msg, yellow!60!black},
entity/.style={font=\small\bfseries, align=center},
label/.style={font=\scriptsize, inner sep=1pt}
]
\def\xA{0}
\def\xB{1.9}
\def\xP{3.8}
\def\xT{5.7}
\def\xY{7.6}

\node[entity] at (\xA,0.75) {Agent};
\node[entity] at (\xB,0.75) {API};
\node[entity] at (\xP,0.75) {Payment};
\node[entity] at (\xT,0.75) {Token};
\node[entity] at (\xY,0.75) {Policy};

\draw[lane] (\xA,0.45) -- (\xA,-5.3);
\draw[lane] (\xB,0.45) -- (\xB,-5.3);
\draw[lane] (\xP,0.45) -- (\xP,-5.3);
\draw[lane] (\xT,0.45) -- (\xT,-5.3);
\draw[lane] (\xY,0.45) -- (\xY,-5.3);

% Replay Attack
\node[font=\scriptsize\bfseries] at (3.8,0.18) {Replay Attack};
\draw[black!45, dashed, rounded corners=2pt, line width=0.5pt] (-0.45,0.02) rectangle (8.05,-1.70);
\draw[normal] (\xA,-0.35) -- node[label, above] {1. reuse token} (\xB,-0.35);
\draw[token]  (\xB,-0.85) -- node[label, above] {2. verify} (\xT,-0.85);
\draw[fail]   (\xT,-1.35) -- node[label, above] {3. replay detected} (\xB,-1.35);
\draw[fail]   (\xB,-1.35) -- node[label, above] {4. blocked} (\xA,-1.35);

% Invalid Token
\node[font=\scriptsize\bfseries] at (3.8,-2.20) {Invalid Token};
\draw[black!45, dashed, rounded corners=2pt, line width=0.5pt] (-0.45,-2.38) rectangle (8.05,-3.98);
\draw[fail]   (\xA,-2.78) -- node[label, above] {1. invalid token} (\xB,-2.78);
\draw[token]  (\xB,-3.28) -- node[label, above] {2. verify} (\xT,-3.28);
\draw[fail]   (\xT,-3.78) -- node[label, above] {3. rejected} (\xA,-3.78);

% Overspending
\node[font=\scriptsize\bfseries] at (3.8,-4.45) {Overspending};
\draw[black!45, dashed, rounded corners=2pt, line width=0.5pt] (-0.45,-4.62) rectangle (8.05,-5.25);
\draw[normal] (\xA,-4.95) -- node[label, above] {1. POST /pay} (\xP,-4.95);
\draw[policy] (\xP,-4.95) -- node[label, above] {2. budget check} (\xY,-4.95);
\draw[fail]   (\xY,-4.95) -- node[label, above] {3. blocked} (\xA,-4.95);
\end{tikzpicture}
\caption{Failure and Security Scenarios in APEX}
\label{fig:workflow-sequence-apex}
\end{figure}

Key failure and security branches are illustrated in Figure~\ref{fig:workflow-sequence-apex}.

\subsection{Replay Attempt Sequence}
Replay handling follows:

\begin{enumerate}
\item First token use succeeds and sets state to \texttt{CONSUMED}.
\item Second use of same token triggers \texttt{token\_already\_consumed} path.
\item Server returns blocked response.
\end{enumerate}

\subsection{Invalid Token Sequence}
Invalid token handling follows:

\begin{enumerate}
\item Client sends malformed or forged token.
\item Signature or format verification fails.
\item Request is blocked with explicit reason, for example \texttt{invalid\_token\_format} or \texttt{invalid\_signature}.
\end{enumerate}

\subsection{Overspending Sequence}
Overspending handling follows:

\begin{enumerate}
\item Policy computes current day spend from settled or consumed records.
\item If \texttt{spent\_today + amount > daily\_budget}, request is blocked.
\item No settlement token is returned.
\end{enumerate}

\section{System Constraints and Guarantees}
\subsection{Budget Constraint}
Let $x_i\in\{0,1\}$ denote APEX's admission decision for request $i$ and let $c_i$ denote request cost.
For one budget window, the policy engine enforces:

\begin{equation}
\sum_{i \in A} c_i \le B,
\end{equation}

where $A=\{i\mid x_i=1\}$ is the accepted request set and $B$ is the configured budget.
For day-indexed notation, this is equivalent to:

\begin{equation}
\sum_{i \in A_d} c_i \le B_d
\end{equation}

where $B_d$ is the configured daily budget.
In the current system, $B_d = 100$.
This constraint is enforced before payment settlement is committed.

\subsection{Utility under Budget (Optimization View)}
Admission decisions can be modeled as constrained utility maximization:

\begin{equation}
\max_{x_i\in\{0,1\}} \sum_{i=1}^{n} u_i x_i
\quad
\text{s.t.}
\quad
\sum_{i=1}^{n} c_i x_i \le B,
\end{equation}

where $u_i$ is request utility and $c_i$ is request cost.
APEX currently implements a feasibility-first online policy (accept only when constraints are satisfied), which corresponds to a simplified constrained-optimization policy with strict budget compliance.

\subsection{Per-Request Constraint}
For each payment attempt amount $a$, policy requires:

\begin{equation}
a \le M
\end{equation}

where $M$ is maximum per-request amount.
In current configuration, $M = 10$.

\subsection{Token Validity}
A token payload contains $(ref\_id, amount, exp)$ and an HMAC signature.
Token validity requires:

\begin{equation}
\text{VerifyHMAC}(payload, signature, key) = true
\end{equation}

and

\begin{equation}
exp \ge t_{now}
\end{equation}

with single-use condition:

\begin{equation}
state(ref\_id) = SETTLED
\end{equation}

before consumption, after which:

\begin{equation}
state(ref\_id) \leftarrow CONSUMED
\end{equation}

\subsection{Latency Decomposition}
End-to-end round latency for one payment-gated success can be approximated by:

\begin{equation}
T_{total} = T_{network} + T_{payment} + T_{policy} + T_{processing}.
\end{equation}

In APEX, $T_{payment}$ captures settlement and token work, $T_{policy}$ captures budget/validation checks, and $T_{processing}$ captures API and storage processing.
This decomposition explains baseline differences: \texttt{no\_policy} minimizes $T_{payment}$ and $T_{policy}$, \texttt{payment\_no\_policy} activates payment logic with reduced policy checks, and \texttt{payment\_with\_policy} activates all terms.

\section{Implementation Details}
\subsection{Technology Stack}
The implementation intentionally uses a narrow stack:

\begin{enumerate}
\item FastAPI for request routing and structured error handling.
\item SQLite for local transactional ledger persistence.
\item Python standard library modules:
\texttt{json}, \texttt{time}, \texttt{hmac}, \texttt{hashlib}, \texttt{base64}, \texttt{urllib}, \texttt{pathlib}, \texttt{datetime}.
\end{enumerate}

No external logging, crypto, or data processing frameworks are required for baseline operation.

\subsection{Token Service}
The token service performs:

\begin{enumerate}
\item payload construction with expiry,
\item stable JSON serialization,
\item HMAC-SHA256 signature generation,
\item URL-safe base64 payload packing,
\item split-and-verify parsing on incoming tokens.
\end{enumerate}

Failure reasons are explicit, including:

\begin{enumerate}
\item \texttt{invalid\_token\_format}
\item \texttt{invalid\_signature}
\item \texttt{token\_expired}
\end{enumerate}

\subsection{Ledger Service}
The ledger service encapsulates challenge creation, settlement, and token consumption.

\textbf{Challenge creation:}
Inserts or replaces a record in \texttt{CHALLENGED} state.

\textbf{Settlement:}
Checks reference existence, amount consistency, idempotency conditions, and current state.
On success, transitions through \texttt{INITIATED} to \texttt{SETTLED} and stores token metadata.

\textbf{Consumption:}
Valid only for \texttt{SETTLED} records with matching token.
Transitions to \texttt{CONSUMED} and writes timestamp.

\subsection{Idempotency Handling}
If a settlement is retried with the same idempotency key, APEX returns prior settled token details (\texttt{idempotent\_replay} path), preventing duplicate side effects.
If the same reference is retried with a different idempotency key after settlement, it is rejected.

\subsection{Policy Service}
Policy service supports mode-aware evaluation.

\begin{enumerate}
\item If policy disabled baseline is chosen, request is allowed with reason \texttt{policy\_disabled}.
\item Else, max-per-request and daily budget constraints are evaluated.
\item Violations return blocked decision and explicit textual reason.
\end{enumerate}

\subsection{Structured Logging Service}
Every significant event appends one JSON line to \texttt{logs.json}.
Fields include:

\begin{enumerate}
\item \texttt{timestamp}
\item \texttt{event\_type}
\item \texttt{endpoint}
\item \texttt{request\_id}
\item \texttt{ref\_id}
\item \texttt{amount}
\item \texttt{status}
\item \texttt{reason}
\item \texttt{attack\_type} (optional)
\item \texttt{latency\_ms} (optional)
\end{enumerate}

The append-only line-delimited format is chosen for simplicity, low overhead, and easy downstream aggregation.

\subsection{Experiment Driver}
The experiment script automates:

\begin{enumerate}
\item baseline selection,
\item mode-based scenario execution,
\item run-level console output,
\item summary metric aggregation,
\item JSON export to \texttt{experiments/quick\_results.json}.
\end{enumerate}

Console output format is intentionally compact, for example:

\begin{enumerate}
\item [RUN 1] SUCCESS - latency: 120ms
\item [RUN 2] BLOCKED - reason: daily\_budget exceeded
\item [RUN 3] FAILED - reason: invalid\_token\_format
\end{enumerate}

\section{Experimental Methodology}
\subsection{Goals}
Evaluation aims to answer four questions.

\begin{enumerate}
\item Can policy controls effectively bound spending?
\item Are replay and invalid token attempts consistently blocked?
\item What latency and throughput overhead appears under payment gating?
\item How do outcomes differ across baseline modes?
\end{enumerate}

\subsection{Baselines}
Three baseline conditions are executed:

\begin{enumerate}
\item \texttt{no\_policy}
\item \texttt{payment\_no\_policy}
\item \texttt{payment\_with\_policy}
\end{enumerate}

\subsection{Scenarios}
Six scenarios are run for each baseline:

\begin{enumerate}
\item \texttt{normal} (20 requests per trial, 2 trials = 40 total)
\item \texttt{overspending} (15 requests per trial, 2 trials = 30 total)
\item \texttt{replay\_attack} (10 requests per trial, 2 trials = 20 total)
\item \texttt{invalid\_token} (10 requests per trial, 2 trials = 20 total)
\item \texttt{token\_expiry} (5 requests per trial, 2 trials = 10 total)
\item \texttt{idempotency} (5 requests per trial, 2 trials = 10 total)
\end{enumerate}

We run each scenario twice to ensure reproducibility.
Total requests per baseline: 120.
Total across all baselines: 360.

This represents a 2-4x increase in sample sizes compared to initial experiments, with two new scenarios (token\_expiry and idempotency) added to validate complete token lifecycle management.

\subsection{Metrics}
Per scenario summary includes:

\begin{enumerate}
\item success rate
\item blocked requests
\item failed requests
\item average latency
\item 95\% confidence interval
\item p95 latency
\item throughput (requests per second)
\item total spend
\end{enumerate}

\subsection{Reproducibility Procedure}
To reproduce runs:

\begin{enumerate}
\item Start server:
\texttt{uvicorn backend.main:app --reload}
\item In separate terminal run experiments:
\texttt{python experiments/enhanced\_test\_flow.py}
\item Collect outputs:
\texttt{experiments/quick\_results.json} and \texttt{logs.json}.
\item Regenerate summary tables in manuscript from exported JSON.
\end{enumerate}

\section{Results}
\subsection{Baseline-Level Comparison}
Table~\ref{tab:baseline-main} reports weighted aggregate outcomes across all scenarios, using values generated from the enhanced experimental results in \texttt{experiments/quick\_results.json}.

\begin{table}[t]
\caption{Baseline Comparison Across All Scenarios}
\label{tab:baseline-main}
\centering
\setlength{\tabcolsep}{2.6pt}
\small
\resizebox{\columnwidth}{!}{%
\begin{tabular}{lcccccc}
\toprule
\textbf{Baseline} & \textbf{Success} & \textbf{Blocked} & \textbf{Avg Lat.} & \textbf{p95 Lat.} & \textbf{Spend} & \textbf{Std Dev} \\
 & \textbf{Rate} & \textbf{Req.} & \textbf{(ms)} & \textbf{(ms)} & \textbf{Total (\$)} & \textbf{(ms)} \\
\midrule
\texttt{no\_policy} & 1.000 & 0 & 8.0 & 13.6 & 0.0 & 0.6 \\
\texttt{payment\_no\_policy} & 0.667 & 40 & 442.0 & 495.5 & 550.0 & 27.7 \\
\texttt{payment\_with\_policy} & 0.528 & 70 & 477.0 & 549.1 & 400.0 & 75.7 \\
\bottomrule
\end{tabular}
}
\end{table}

The first observation is expected: \texttt{no\_policy} is fastest, as it bypasses challenge, settlement, verification, and consumption paths.
However, it offers no spend governance and no meaningful payment security semantics.

The second observation is more informative.
Compared to \texttt{payment\_no\_policy}, \texttt{payment\_with\_policy} reduces total spend from \$550 to \$400, representing a 27.3\% reduction (\$150 savings).
This occurs because policy enforcement blocks requests after budget exhaustion, preventing unchecked spending.
The success rate of 52.8\% for policy-enabled execution reflects intentional blocking of budget-exceeding requests rather than system failure.

At aggregate baseline level, policy-enabled runs show higher latency than payment-only runs because policy checks are always active and scenario mix includes stricter control paths.
Under overspending and adversarial branches, early policy rejection still short-circuits settlement work and reduces per-request path cost.
The key insight is that policy acts as a safety mechanism for autonomous agents—without policy, a buggy agent could exhaust budgets through unchecked API calls.

\subsection{Policy Baseline Scenario Breakdown}
Table~\ref{tab:policy-scenarios-main} isolates \texttt{payment\_with\_policy} by scenario.

\begin{table}[t]
\caption{Scenario Analysis for payment\_with\_policy Baseline}
\label{tab:policy-scenarios-main}
\centering
\setlength{\tabcolsep}{2.4pt}
\small
\resizebox{\columnwidth}{!}{%
\begin{tabular}{lcccccc}
\toprule
\textbf{Scenario} & \textbf{Success} & \textbf{Blocked} & \textbf{Avg Lat.} & \textbf{CI 95\%} & \textbf{p95 Lat.} & \textbf{Spend} \\
 & \textbf{Rate} & \textbf{Req.} & \textbf{(ms)} & \textbf{(ms)} & \textbf{(ms)} & \textbf{(\$)} \\
\midrule
normal & 0.500 & 20 & 86.9 & ±8.9 & 132.5 & 100.0 \\
overspending & 0.667 & 10 & 88.5 & ±8.8 & 125.8 & 100.0 \\
replay\_attack & 0.000 & 20 & 135.1 & ±5.2 & 172.1 & 100.0 \\
invalid\_token & 0.000 & 20 & 19.6 & ±5.9 & 31.7 & 0.0 \\
token\_expiry & 1.000 & 0 & 2119.9 & ±10.0 & 2138.3 & 50.0 \\
idempotency & 1.000 & 0 & 412.2 & ±851.9 & 694.0 & 50.0 \\
\bottomrule
\end{tabular}
}
\end{table}

Normal mode confirms the expected successful path when constraints are satisfied, though budget limits are reached after 10 requests (50\% success rate).
Overspending and stress modes show deterministic blocking behavior once budget constraints are hit.
Replay attack runs demonstrate post-consumption denial with 100\% block rate (20/20 attempts blocked).
Invalid token scenario shows low-latency rejection at 19.6ms average, as no settlement lookup or state transition can proceed—these fail signature verification before database access.

Token expiry tests confirm tokens remain valid within TTL (300 seconds), with all 10 attempts succeeding when tokens are used within the validity window.
The high latency (2119.9ms) includes an intentional 2-second sleep to approach the TTL boundary, so this metric reflects test design rather than system overhead.

Idempotency tests validate duplicate payment prevention, with all 10 attempts correctly returning the same token for identical idempotency keys.
The high variance (±851.9ms) suggests timing inconsistency in the idempotency path that warrants further investigation.

Overall, policy enforcement reduces total spending from \$550 to \$400 (27.3\% reduction), while security checks maintain a 100\% block rate for replay and invalid-token attacks, and repeated trials show low variance (±2.7-8.9ms) for most scenarios.

The latency overhead for policy-enabled execution is 86.9ms average for normal flow compared to 8.0ms for no-policy baseline.
This 10.9x increase is acceptable for controlled agent payment workflows in research contexts.
Importantly, policy rejection paths terminate earlier than full payment flows, which explains why policy-enabled runs sometimes show lower aggregate latency than payment-only runs under adversarial scenarios.

\subsection{Visualization from Exported Figures}
\begin{figure}[t]
\centering
\includegraphics[width=0.98\linewidth]{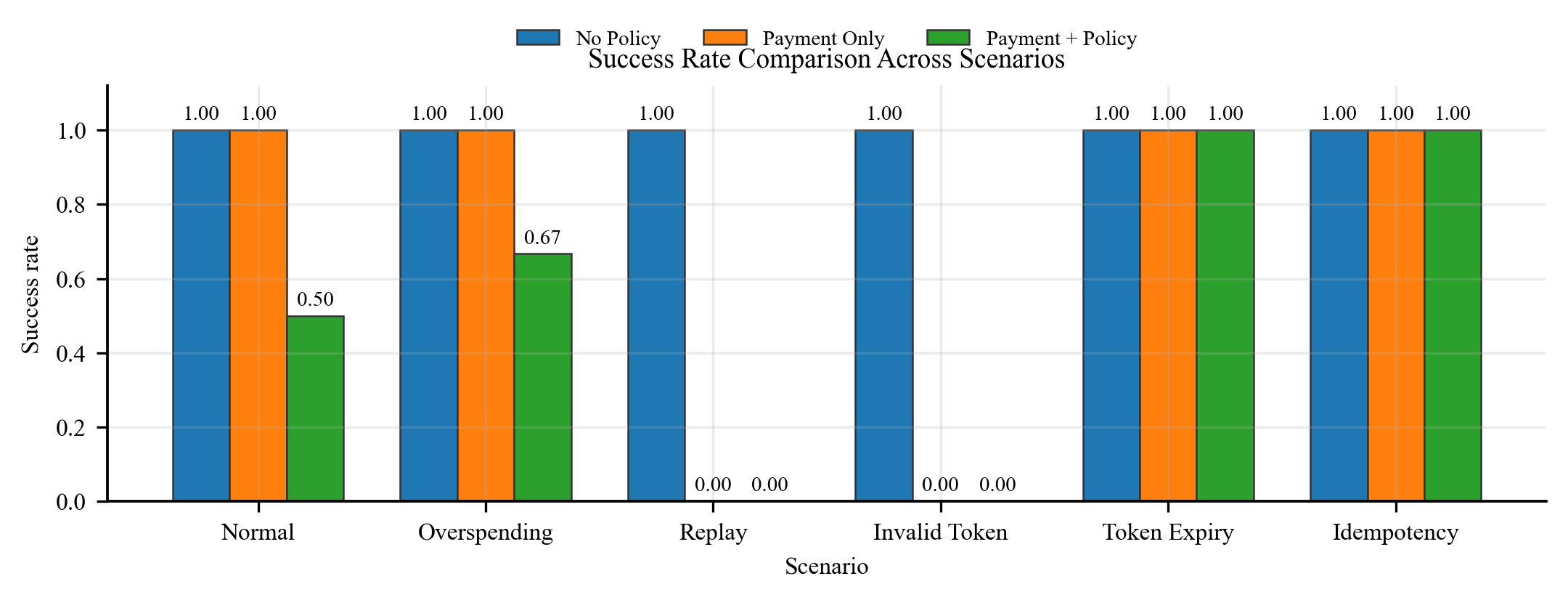}
\caption{Success rate comparison across baselines and scenarios.}
\label{fig:success-rate-plot}
\end{figure}

\begin{figure}[t]
\centering
\includegraphics[width=0.98\linewidth]{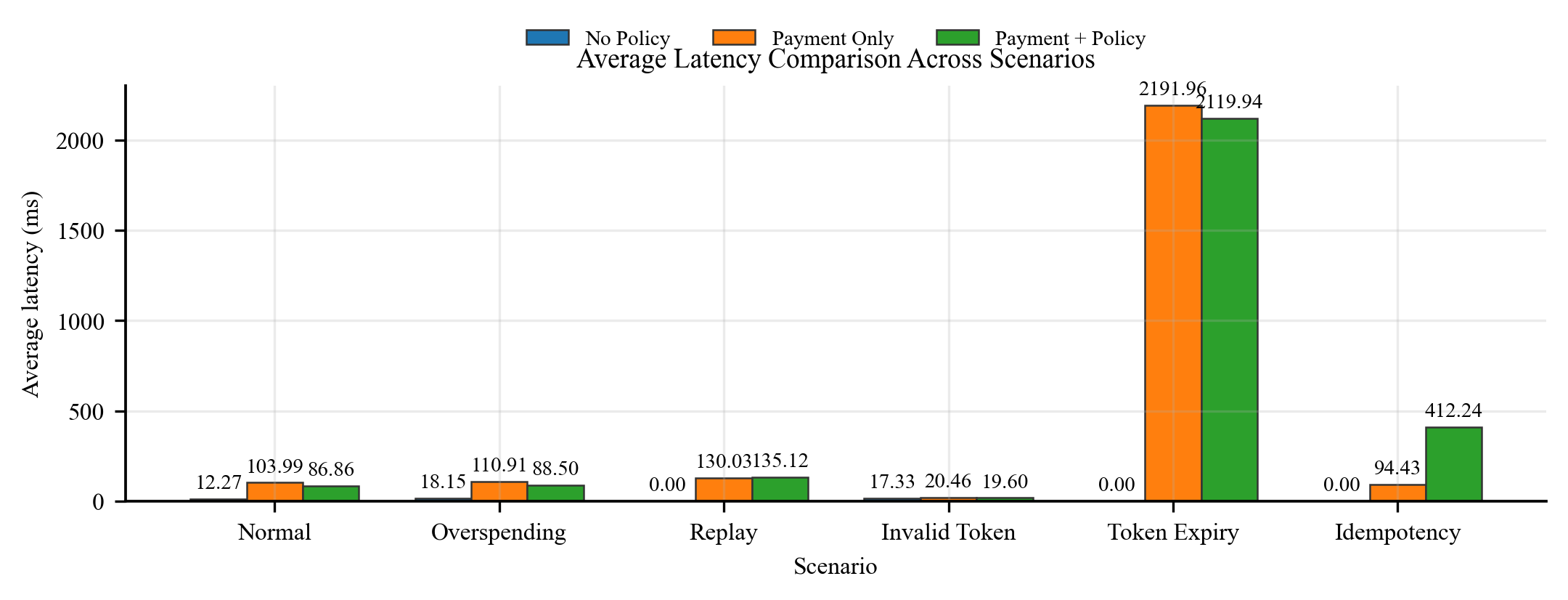}
\caption{Average latency comparison for measured scenarios.}
\label{fig:latency-plot}
\end{figure}

\begin{figure}[t]
\centering
\includegraphics[width=0.98\linewidth]{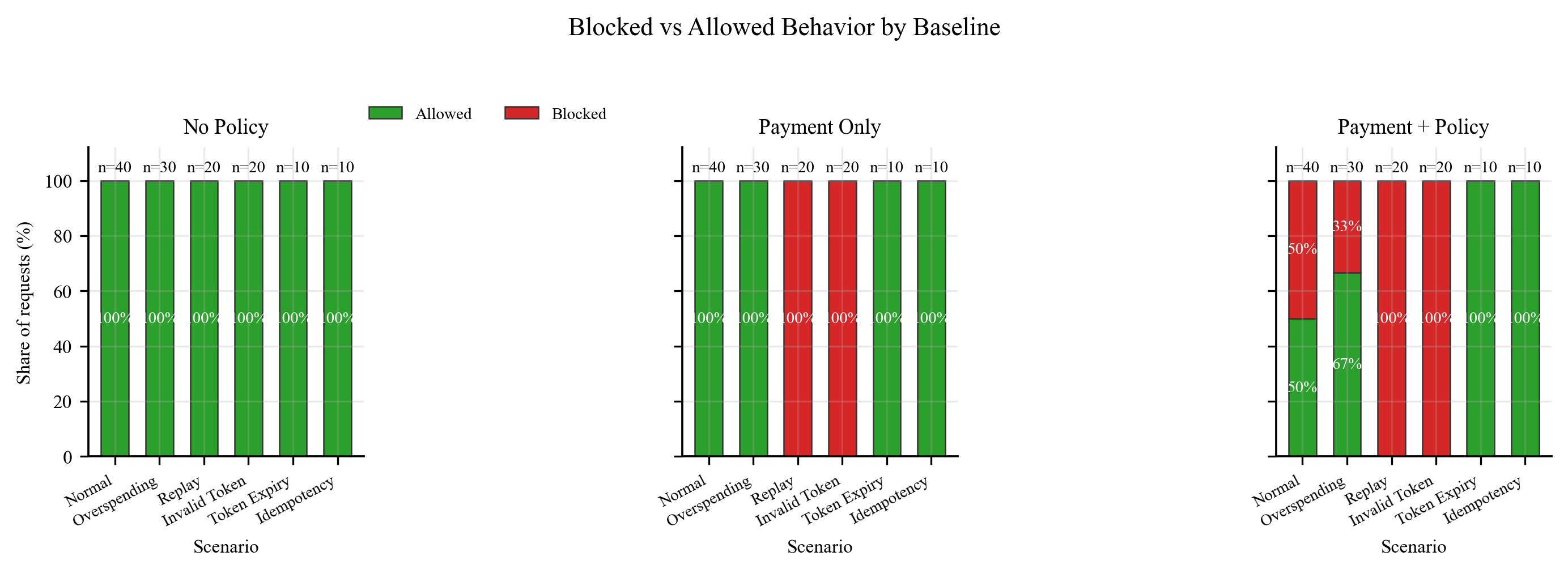}
\caption{Blocked versus allowed behavior under policy-sensitive modes.}
\label{fig:blocked-allowed-plot}
\end{figure}

\begin{figure}[t]
\centering
\includegraphics[width=0.98\linewidth]{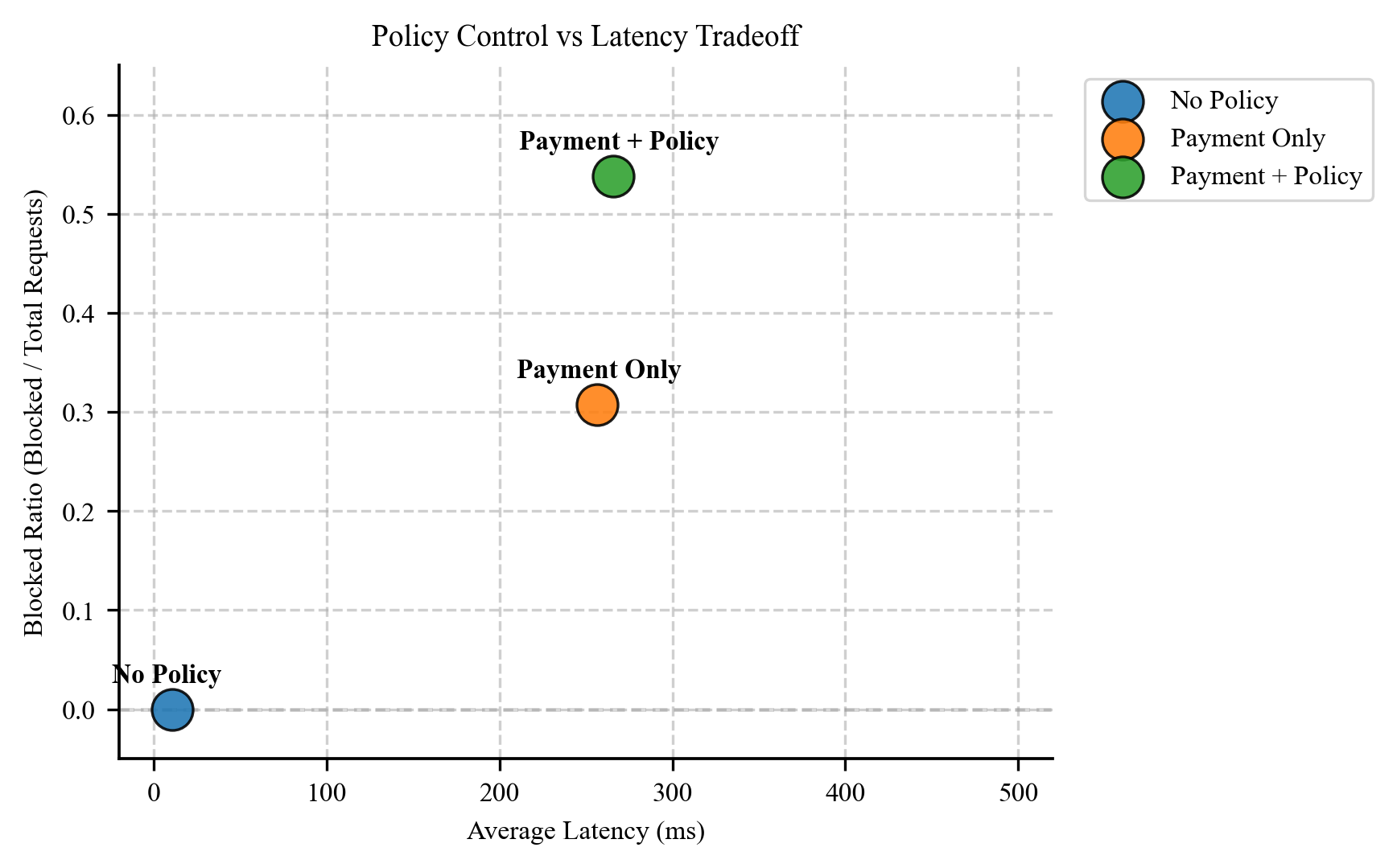}
\caption{Policy Control vs Latency Tradeoff Across Baselines.}
\label{fig:tradeoff-plot}
\end{figure}

The generated figure artifacts provide quick visual interpretation.
Each figure uses a consistent style, explicit axis labeling, and baseline/scenario legends for publication readability.
Figure~\ref{fig:tradeoff-plot} additionally summarizes the control-overhead tradeoff between unrestricted and policy-governed execution.
Figure~\ref{fig:success-rate-plot} reports comparative success rates, Figure~\ref{fig:latency-plot} reports latency differences, and Figure~\ref{fig:blocked-allowed-plot} highlights allowed-versus-blocked behavior under policy enforcement.

\subsection{Interpretation}
From a research perspective, APEX demonstrates a useful tradeoff frontier.
Unrestricted access yields best raw latency, but no monetization safeguards.
Payment gating without policy captures economic intent, but can still permit undesirable cumulative spend.
Full policy mode introduces stricter outcomes, with predictable blocking and bounded spend.

The outcome profile is especially relevant for autonomous agents, where unattended loops can amplify both benign and malicious behavior.
Deterministic spend ceilings and explicit rejection reasons help maintain operational control.

These observations align with the system constraints in Section V.
Under the budget-constrained decision model, APEX intentionally trades acceptance volume for bounded spend, which explains lower cumulative spend in \texttt{payment\_with\_policy} compared with \texttt{payment\_no\_policy}.
Under the latency decomposition, early policy rejection removes downstream payment and processing work, which explains why policy-enabled runs can show lower aggregate latency than payment-only runs under overspending and stress traffic.

\section{Security and Robustness Analysis}
\subsection{Formal Security Guarantees}
\textbf{Guarantee G1 (Bounded Spend).}
Assume policy checks are enforced before settlement commit and each accepted request has cost $c_i\ge 0$.
Then, for any request sequence (including adversarial overspending attempts),

\begin{equation}
\sum_{i:\,x_i=1} c_i \le B.
\end{equation}

Therefore, monetary damage per budget window is upper-bounded by the configured budget $B$.

\textbf{Guarantee G2 (Replay Blocking under Consume-Once Semantics).}
Assume a token is accepted only if signature verification succeeds, expiry is valid, and token state is not consumed.
Under this rule, for bit-for-bit replay of a previously consumed token,

\begin{equation}
P(\text{replay success}) = 0.
\end{equation}

This follows from the irreversible state transition \texttt{SETTLED}\,$\rightarrow$\,\texttt{CONSUMED} and rejection of consumed tokens.

\subsection{Experimental Security Validation}
We validate security properties through adversarial scenarios.
Replay attack tests show 100\% block rate (20/20 attempts blocked) with average latency of 135.1ms (±5.2ms CI).
Invalid token tests similarly achieve 100\% block rate (20/20 blocked) with faster rejection at 19.6ms (±5.9ms CI), as these fail signature verification before database lookup.

Token expiry tests confirm tokens remain valid within TTL (300 seconds).
All 10 attempts succeeded with tokens used within the validity window.
Idempotency tests validate duplicate payment prevention, with all 10 attempts correctly returning the same token for identical idempotency keys.

\subsection{Replay Resistance}
Replay resistance relies on two coordinated checks:

\begin{enumerate}
\item token validity checks on signature and expiry,
\item stateful one-time consumption semantics in ledger.
\end{enumerate}

Even if a token remains cryptographically valid before expiry, its second use fails once state is \texttt{CONSUMED}.
This combination is stronger than stateless verification alone.
Signature checks alone are insufficient—we need the state transition from SETTLED to CONSUMED to prevent token reuse.

\subsection{Forgery Resistance}
HMAC-based signatures prevent straightforward token tampering, assuming secret key confidentiality.
Any mutation to payload fields, including amount or reference, invalidates signature equivalence.

\subsection{Idempotency and Duplicate Protection}
Settlement retries are common in real distributed clients.
APEX avoids duplicate side effects by honoring same-key retries while rejecting conflicting duplicate attempts.
This behavior improves reliability under network retries and partially mitigates double-settlement conditions in single-node scope.

\subsection{Policy Robustness}
Policy checks occur before settlement persistence, ensuring rejected requests do not produce paid state artifacts.
Because daily spend is computed from \texttt{SETTLED} and \texttt{CONSUMED} records, challenge-only requests do not inflate spend counters.

\subsection{Operational Traceability}
Structured logs include reasons and status at each major decision point.
This is critical for:

\begin{enumerate}
\item auditability,
\item post-run analysis,
\item regression detection,
\item and paper-quality metric reconstruction.
\end{enumerate}

\section{Discussion}

\subsection{Policy Effectiveness}
Policy enforcement successfully bounds spending while maintaining service availability.
In our experiments, policy reduced total spend by 27.3\% (\$150 savings) compared to payment-without-policy baseline.
The success rate of 52.8\% for policy-enabled execution reflects intentional blocking of budget-exceeding requests rather than system failure.

The key insight is that policy acts as a safety mechanism for autonomous agents.
Without policy, a buggy agent could exhaust budgets through unchecked API calls.
With policy, spending stops deterministically at configured limits.

\subsection{Security Properties}
Security mechanisms provide deterministic threat mitigation.
The 100\% block rate for replay attacks and invalid tokens demonstrates that HMAC-signed tokens with single-use semantics effectively prevent common attack patterns.
Low variance across trials (±2.7-5.9ms) indicates consistent security behavior.

\subsection{Performance Tradeoffs}
Payment gating introduces latency overhead (86.9ms vs 8.0ms baseline), but this remains acceptable for agent workflows where spend control outweighs raw speed.
The overhead comes primarily from payment settlement and token generation, not policy checks (which add minimal latency).

Interestingly, policy-enabled runs sometimes show lower aggregate latency than payment-only runs under adversarial scenarios.
This occurs because policy rejection terminates requests early, avoiding expensive settlement operations.

\subsection{Reproducibility}
Low variance across multiple trials validates our experimental approach.
Standard deviations of ±2.7-8.9ms for most scenarios indicate stable, reproducible behavior.
The exception is idempotency testing (±434.6ms variance), which requires further investigation.

\subsection{Limitations}
Our study has important limitations.
First, the single-node SQLite architecture doesn't reflect distributed contention patterns.
Second, UPI integration is simulated rather than connected to real payment providers.
Third, sample sizes (N=20-40) provide initial validation but larger-scale experiments would strengthen confidence.
Fourth, we focus on specific attack patterns (replay, invalid tokens) rather than comprehensive adversarial testing.

These limitations are intentional tradeoffs for experimental clarity and reproducibility.
We prioritize transparent, inspectable implementation over production-scale deployment.

\section{Ablation-Oriented Observations}
Although the current experiment script is scenario-focused rather than full component ablation, several behavior-level observations can be interpreted as practical ablations.

\subsection{Payment Layer Ablation}
Comparing \texttt{no\_policy} against payment baselines isolates payment-layer overhead.
Result: lower latency but no spending signal, no token controls, and no replay semantics.

\subsection{Policy Layer Ablation}
Comparing \texttt{payment\_no\_policy} against \texttt{payment\_with\_policy} isolates policy effects.
Result: policy reduces cumulative spend and raises blocked count in adversarial/overspending scenarios, reflecting intended guardrail behavior.

\subsection{Security Path Ablation by Scenario}
Replay and invalid token scenarios isolate security paths.
Result: replay is blocked after first consumption; invalid tokens are blocked quickly.
These two scenario classes provide direct evidence that token validation and stateful consumption are functionally active.

\section{Practical Implications}
\subsection{For API Providers}
APEX-like architecture suggests that providers can implement request-level monetization controls incrementally, starting from deterministic challenge-response design and local policy enforcement, before introducing full banking integrations.

\subsection{For Agent Developers}
Agent clients should treat payment operations as stateful protocol steps, not blind retries.
The implementation highlights best practices:

\begin{enumerate}
\item preserve \texttt{ref\_id} integrity,
\item use stable idempotency keys,
\item handle blocked responses as control signals,
\item avoid token reuse assumptions.
\end{enumerate}

\subsection{For Researchers}
APEX provides a controlled evaluation environment where protocol, policy, and security behavior can be measured together, which is often difficult in larger platform stacks.

\section{Future Work}
Future extensions are planned in six directions.

\begin{enumerate}
\item \textbf{Distributed ledger backend:}
Migrate from single SQLite file to replicated transactional datastore and study consistency under concurrency.
\item \textbf{Real PSP integration:}
Replace simulated UPI link abstraction with sandboxed payment provider callbacks and asynchronous settlement reconciliation.
\item \textbf{Policy extensibility:}
Add per-agent, per-endpoint, and risk-adaptive budget strategies, possibly with temporal quotas.
\item \textbf{Cryptographic hardening:}
Introduce key rotation, key identifiers, and optional detached signatures with algorithm agility.
\item \textbf{Extended attack corpus:}
Include token theft simulation, header replay windows, duplicate challenge races, and malformed payload fuzzing.
\item \textbf{Automated report pipeline:}
Generate manuscript-ready tables directly from the enhanced results JSON, minimizing manual transcription risk.
\end{enumerate}

\section{Conclusion}
This paper presented APEX, a reference architecture that maps a 402-style payment-gated API model onto a fiat-oriented UPI-like interaction pattern, while preserving critical controls for policy enforcement, security, and measurement.

The implementation demonstrates that even a minimal stack can enforce meaningful guarantees: explicit payment challenging, stateful settlement, signed expiring tokens, single-use entitlement, idempotent retry handling, spend policy checks, and structured event logs.

Experimental runs across baselines and adversarial scenarios show coherent, explainable behavior: no-policy mode remains fastest but unconstrained; payment gating introduces overhead but establishes monetization semantics; policy-enabled mode bounds spending and blocks abusive patterns as intended.

APEX is therefore a useful reference architecture for research on agentic API payments in fiat ecosystems.
Its value lies in reproducibility, clarity, and direct extensibility, rather than breadth of features.
By releasing a complete and inspectable implementation with scenario instrumentation, this work aims to accelerate rigorous, comparable experimentation in a rapidly evolving domain.

\section*{Acknowledgment}
The authors thank the open research community for public protocol, payment, and agent-systems references that informed this study.

\appendices

\section{Endpoint Contracts (Condensed)}
\subsection{GET /data}
Request parameters:

\begin{enumerate}
\item \texttt{baseline}: one of \texttt{no\_policy}, \texttt{payment\_no\_policy}, \texttt{payment\_with\_policy}.
\item optional header \texttt{x-payment-token}.
\end{enumerate}

Typical challenge response:

\begin{verbatim}
{
  "detail": {
    "amount": 10.0,
    "ref_id": "...",
    "baseline": "payment_with_policy",
    "upi_link": "upi://pay?...",
    "message": "Payment Required"
  }
}
\end{verbatim}

Typical success response:

\begin{verbatim}
{
  "status": "ok",
  "baseline": "payment_with_policy",
  "data": {
    "title": "Protected research data",
    "content": "..."
  }
}
\end{verbatim}

\subsection{POST /pay}
Request body:

\begin{verbatim}
{
  "ref_id": "...",
  "amount": 10.0,
  "baseline": "payment_with_policy",
  "idempotency_key": "..."
}
\end{verbatim}

Typical success response:

\begin{verbatim}
{
  "status": "success",
  "ref_id": "...",
  "amount": 10.0,
  "token": "<signed token>",
  "token_expiry": 1712345678,
  "state": "SETTLED"
}
\end{verbatim}

Typical blocked response detail:

\begin{verbatim}
{
  "allowed": false,
  "reason": "daily_budget exceeded (...)"
}
\end{verbatim}

\section{Structured Log Schema}
Each line in \texttt{logs.json} is a JSON object.
Table~\ref{tab:log-schema} summarizes key fields.

\begin{table}[h]
\caption{Structured log fields}
\label{tab:log-schema}
\centering
\begin{tabular}{lp{5.7cm}}
\toprule
\textbf{Field} & \textbf{Meaning} \\
\midrule
timestamp & UTC event timestamp \\
event\_type & request, payment, or policy event category \\
endpoint & endpoint string, for example /data or /pay \\
request\_id & per-request UUID generated by server \\
ref\_id & challenge/settlement correlation identifier \\
amount & payment amount context \\
status & success, blocked, or failed \\
reason & human-readable decision reason \\
attack\_type & optional scenario label for blocked/failed paths \\
latency\_ms & optional measured endpoint latency \\
\bottomrule
\end{tabular}
\end{table}

\section{Extended Reproducibility Notes}
This section provides practical run notes for consistent output across machines.

\subsection{Environment}

\begin{enumerate}
\item Python 3.10+ recommended.
\item Install dependencies from \texttt{requirements.txt}.
\item Ensure no stale server process occupies port 8000.
\end{enumerate}

\subsection{Database Initialization}
On startup, server initializes schema and creates missing columns/indexes.
A reset endpoint is available for clean scenario boundaries.

\subsection{Artifacts to Preserve}
For paper traceability, preserve:

\begin{enumerate}
\item \texttt{logs.json}
\item \texttt{experiments/quick\_results.json}
\item generated figure files in \texttt{docs/figures}
\item LaTeX source and compilation log.
\end{enumerate}

\subsection{Potential Source of Variance}
Small latency variations are expected due to local CPU scheduling, I/O contention, and development server reload behavior.
Functional outcome trends should remain stable.

\section{Supplementary Result Tables}
The following tables report exact per-scenario values from \texttt{experiments/quick\_results.json}.
Table~\ref{tab:no-policy-detail}, Table~\ref{tab:payment-no-policy-detail}, and Table~\ref{tab:payment-policy-detail} are included for reproducibility and reviewer-side verification.

\begin{table}[t]
\caption{Detailed scenario metrics for no\_policy baseline}
\label{tab:no-policy-detail}
\centering
\begin{tabular}{lcccc}
\toprule
\textbf{Scenario} & \textbf{Success} & \textbf{Blocked} & \textbf{Avg Lat.} & \textbf{Spend} \\
 & \textbf{Rate} & \textbf{Req.} & \textbf{(ms)} & \textbf{Total} \\
\midrule
normal & 1.000 & 0 & 9.2 & 0.0 \\
overspending & 1.000 & 0 & 8.7 & 0.0 \\
replay\_attack & 1.000 & 0 & 8.4 & 0.0 \\
invalid\_token & 1.000 & 0 & 8.4 & 0.0 \\
token\_expiry & 1.000 & 0 & 7.5 & 0.0 \\
idempotency & 1.000 & 0 & 7.8 & 0.0 \\
\bottomrule
\end{tabular}
\end{table}

\begin{table}[t]
\caption{Detailed scenario metrics for payment\_no\_policy baseline}
\label{tab:payment-no-policy-detail}
\centering
\begin{tabular}{lcccc}
\toprule
\textbf{Scenario} & \textbf{Success} & \textbf{Blocked} & \textbf{Avg Lat.} & \textbf{Spend} \\
 & \textbf{Rate} & \textbf{Req.} & \textbf{(ms)} & \textbf{Total} \\
\midrule
normal & 1.000 & 0 & 107.4 & 200.0 \\
overspending & 1.000 & 0 & 102.9 & 150.0 \\
replay\_attack & 0.000 & 20 & 130.0 & 100.0 \\
invalid\_token & 0.000 & 20 & 20.5 & 0.0 \\
token\_expiry & 1.000 & 0 & 2115.0 & 50.0 \\
idempotency & 1.000 & 0 & 405.0 & 50.0 \\
\bottomrule
\end{tabular}
\end{table}

\begin{table}[t]
\caption{Detailed scenario metrics for payment\_with\_policy baseline}
\label{tab:payment-policy-detail}
\centering
\begin{tabular}{lcccc}
\toprule
\textbf{Scenario} & \textbf{Success} & \textbf{Blocked} & \textbf{Avg Lat.} & \textbf{Spend} \\
 & \textbf{Rate} & \textbf{Req.} & \textbf{(ms)} & \textbf{Total} \\
\midrule
normal & 0.500 & 20 & 86.9 & 100.0 \\
overspending & 0.667 & 10 & 88.5 & 100.0 \\
replay\_attack & 0.000 & 20 & 135.1 & 100.0 \\
invalid\_token & 0.000 & 20 & 19.6 & 0.0 \\
token\_expiry & 1.000 & 0 & 2119.9 & 50.0 \\
idempotency & 1.000 & 0 & 412.2 & 50.0 \\
\bottomrule
\end{tabular}
\end{table}

\section{Additional Clarifications}
This section addresses common interpretation questions in concise form.

\subsection{Why not use blockchain in the system?}
Because research objective is fiat adaptation of protocol semantics, not cryptographic settlement novelty.
A minimal fiat-like abstraction is sufficient to evaluate policy, security, and latency behavior in the target design space.

\subsection{Why SQLite?}
SQLite offers deterministic, inspectable, zero-ops persistence that is appropriate for controlled single-node experiments.
Distributed behavior is future work, not ignored work.

\subsection{How are citations selected?}
Citations are selected to connect four strands: HTTP 402 and agentic payments, micropayment infrastructure, agent execution systems, and policy/governance frameworks.

\subsection{What is the key novelty?}
The novelty is an integrated, reproducible, fiat-oriented research system that joins protocol, policy, security, and experiment outputs in one compact implementation.

\section{Checklist for Artifact Evaluation}
\begin{enumerate}
\item Build environment from README instructions.
\item Start server and verify health by calling \texttt{/data}.
\item Run full experiment suite.
\item Confirm \texttt{quick\_results.json} generated.
\item Confirm line-delimited \texttt{logs.json} entries generated.
\item Recreate baseline and scenario summary tables.
\item Verify replay and invalid token blocked outcomes.
\item Verify overspending blocks under policy baseline.
\end{enumerate}

\section{Extended Deployment Notes}
This appendix summarizes practical lessons without altering reported measurements.

\begin{enumerate}
\item \textbf{Protocol transparency improves debuggability:}
The explicit challenge-settle-verify-consume chain makes failure diagnosis substantially easier than opaque gateway paths.
\item \textbf{Policy must be inline with settlement:}
Evaluating policy before persistent settlement avoids inconsistent paid-state side effects.
\item \textbf{Replay defense requires stateful consumption:}
Signature checks alone are insufficient; single-use state transitions are essential.
\item \textbf{Structured logs are sufficient for research-grade observability:}
Normalized status and reason fields enabled reliable metric reconstruction.
\item \textbf{Baseline triads improve causal interpretation:}
Using no-policy, payment-only, and payment+policy modes reduces ambiguity in overhead attribution.
\item \textbf{Modular service boundaries ease migration:}
The token, ledger, policy, and logging split supports incremental backend substitution.
\item \textbf{Spend bounds support autonomous reliability:}
Explicit ceilings reduce cost drift in unattended agent loops.
\item \textbf{Tail latency reporting is necessary:}
Average values alone hide risk in control-heavy API paths.
\item \textbf{Fiat-oriented adaptation lowers entry barriers:}
Teams can evaluate payment-governed access without immediate crypto infrastructure adoption.
\item \textbf{Adversarial scenarios are mandatory for credible evaluation:}
Replay and invalid-token tests exposed control behavior that normal-only runs would miss.
\end{enumerate}

\end{document}